\documentclass[preprint,showpacs,preprintnumbers,amsmath,amssymb]{revtex4}

\usepackage{graphicx}% Include figure files
\usepackage{dcolumn}% Align table columns on decimal point
\usepackage{bm}% bold math

\begin{document}

%\preprint{APS/123-QED}

\title{A catastrophe model for fast magnetic reconnection onset}
%\title{Manuscript Title:\\with Forced Linebreak}% Force line breaks with \\

\author{P.~A.~Cassak}
%\altaffiliation[Also at ]{Physics Department, XYZ University.}
\author{M.~A.~Shay}%
\author{J.~F.~Drake}%
\affiliation{%
University of Maryland, College Park, MD 20742
}%

%\affiliation{

\date{\today}% It is always \today, today,

\begin{abstract}
A catastrophe model for the onset of fast magnetic reconnection is
presented that suggests why plasma systems with magnetic free energy
remain apparently stable for long times and then suddenly 
release their energy. For a given set of plasma parameters there are
generally two stable reconnection solutions: a slow (Sweet-Parker)
solution and a fast (Alfv\'enic) Hall reconnection
solution. Below a critical resistivity the slow solution disappears
and fast reconnection dominates.  Scaling arguments predicting the two
solutions and the critical resistivity are confirmed with two-fluid
simulations.
\end{abstract}

\maketitle

Explosive events in plasmas, such as solar eruptions and the sawtooth
crash in laboratory fusion devices, are driven by magnetic
reconnection.  Understanding the mechanism facilitating fast
reconnection in high temperature plasma systems has been a
long-standing challenge. Sweet-Parker (SP) reconnection 
\cite{Sweet58,Parker57} is far too slow to explain observations and 
Petschek reconnection requires the invocation of anomalous 
resistivity, a phenomenon that is at best only poorly
understood. A new paradigm has emerged in recent years in which
dispersive whistler and kinetic Alfv\'en waves facilitate fast
reconnection by setting up the open Petschek configuration
\cite{Shay99,Birn01,Rogers01}. Magnetospheric satellite observations
\cite{Oieroset01} and recent laboratory experiments support this new 
paradigm \cite{Ren05}. 

It is not sufficient, however, to merely explain how fast reconnection
can occur. If reconnection were always fast, magnetic stress could
never build up in plasma systems such as the solar corona and the
explosive release of magnetic energy seen in nature and the laboratory
would never occur. It is critical, therefore, to explain why fast
reconnection does not always take place. We show that there are
generally two reconnection solutions for a given set of parameters:
slow reconnection as predicted by Sweet and Parker; and fast
collisionless reconnection facilitated by coupling to dispersive waves
in the dissipation region (Hall reconnection). Below a critical
resistivity the slow solution disappears. The emerging picture,
therefore, is that slow reconnection can dominate the dynamics of a
system for long periods of time but the resulting rate of
reconnection is so slow that external forces can continue build up
magnetic stresses. When the resistivity $\eta$ drops below a critical 
value (or the available free energy crosses a threshold) the system 
abruptly transitions to fast reconnection and is manifest as a magnetic
explosion. Such a model complements earlier ideas that the onset
of solar flares, for example, results from the loss of MHD
equilibrium \cite{Hesse86,Lin03} or more complex ``breakout''
models \cite{Antiochos99}.

A rather simple argument can be made to motivate why magnetic
reconnection is bistable, {\it i.e.}, has two solutions for a given
set of parameters.  The SP solution is valid provided the half width 
of the current layer $\delta$ exceeds the relevant kinetic scale lengths 
\cite{Biskamp00},
\begin{equation}
\frac{\delta}{L} = \sqrt{\frac{\eta c^{2}}{4 \pi c_{A} L}} > 
\frac{d_{i}}{L},\frac{\rho_{s}}{L},     \label{sfcond}
\end{equation}
where $L$ is the half length of the SP current sheet, $d_{i} = c/ \omega_{pi}$
is the ion inertial length, $\rho_{s}$ is the ion Larmor radius, $\omega_{pi}$ 
is the ion plasma frequency, and $c_{A}$ is the Alfv\'en speed.
The Alfv\'en speed is to be evaluated immediately upstream of the current 
layer.  Therefore, if the system is undergoing SP reconnection and the 
resistivity is lowered, SP reconnection will continue as long as 
Eq.~(\ref{sfcond}) is satisfied. 

Conversely, fast reconnection is valid provided the dispersive (whistler
or kinetic Alfv\'en) waves that drive kinetic reconnection \cite{Rogers01}
are not dissipated.  We restrict our discussion to whistler waves, 
generated by the Hall term.  The dispersion relation for resistive whistler 
waves is $\omega = k^{2} c_{A} d_{i} - i k^{2} \eta c^{2}/4 \pi$.  Since 
both terms scale like $k^{2}$, dissipation can only be neglected if it 
is small enough at all spatial scales, that is
\begin{equation}
\frac{\eta c^{2}}{4 \pi} \ll c_{A} d_{i}.  \label{collfreq}
\end{equation}
This can also be written as $\nu_{ei} \ll \Omega_{ce}$, where
$\nu_{ei} = \eta n e^{2} / m_{e}$ is the electron-ion collision
frequency and $\Omega_{ce} = e B / m_{e} c$ is the electron cyclotron
frequency, a condition which is typically easily satisfied in nature.
Therefore, if the system is undergoing Hall reconnection and
the resistivity is increased, it will stay in the Hall configuration as 
long as Eq.~(\ref{collfreq}) is satisfied.  If the resistivity is an 
intermediate value such that both Eqs.~(\ref{sfcond}) and (\ref{collfreq}) 
are satisfied, then either solution is accessible and the system is bistable.

We now present estimates of the slow-to-fast
($\eta_{sf}$) and fast-to-slow ($\eta_{fs}$) resistive transition
boundaries of the bistable regime.  We estimate $\eta_{sf}$ by setting
the left and right hand sides of Eq.~(\ref{sfcond}) equal using
$d_{i}$ as the relevant kinetic scale length for Hall physics:
\begin{equation}
\eta_{sf} \frac{c^{2}}{4 \pi} \sim \frac{c_{A} d_{i}^{2}}{L}.   \label{etasf}
\end{equation}
To estimate $\eta_{fs}$, we perform a Sweet-Parker type scaling analysis 
\cite{Parker57} that is more precise than the argument used to motivate 
bistability in Eq.~(\ref{collfreq}).  Resistive effects are negligible if 
the outward magnetic diffusion across the electron current sheet, $\eta c^2 
/ 4 \pi \delta^2$, is less than the inward convection, $v_{in} / 
\delta$, where $v_{in}$ is the flow speed into the electron current layer.  
For Hall reconnection, numerical simulations \cite{Shay99} have 
shown that $\delta$ scales like the electron inertial length $d_{e} = c/ 
\omega_{pe}$, where $\omega_{pe} = \sqrt{4 \pi n e^{2} / m_{e}}$ is the 
electron plasma frequency, and the inflow speed scales like $v_{in} \sim 
0.1 c_{Ae}$, where $c_{Ae}$ is the electron Alfv\'en speed based on the 
magnetic field immediately upstream of the electron current layer.  The 
critical resistivity $\eta_{fs}$ is found by equating the two: 
\[
\eta_{fs} \frac{c^{2}}{4 \pi d_{e}^{2}} \sim \frac{v_{in}}
{d_{e}} \sim 0.1 \frac{c_{Ae}}{d_{e}},
\]
or, using $c_{Ae} d_{e} = c_{A} d_{i}$,
\begin{equation}
\eta_{fs} \frac{c^{2}}{4 \pi} \sim 0.1 c_{A} d_{i},   \label{etafs}
\end{equation}
where $c_{A}$ is evaluated upstream of the {\it electron} current layer.
This is consistent with Eq.~(\ref{collfreq}), but more precise since
the geometry of the layer is included.  Note that $\eta_{fs}$ is
independent of system size and electron mass and is enormous for most
physical systems.  Equation (\ref{etafs}) suggests that once Hall
reconnection onsets, resistive effects are unlikely to influence the
dynamics.  The ratio of Eqs.~(\ref{etasf}) and (\ref{etafs}) gives
$\eta_{sf} / \eta_{fs} \sim 10 d_{i} / L \ll 1$, which is small
because $d_{i} \ll L$ for most systems of physical interest. Thus,
bistabilty is present over an enormous range of resistivity.

The predictions of this model are amenable to tests using numerical
simulations.  We use the two-fluid code, f3d, a massively parallel code
described elsewhere \cite{Shay04}, to perform two-dimensional
simulations in a slab geometry of size $L_{x} \times L_{y}$.  The
initial equilibrium is two Harris sheets, ${\bf B} = {\bf \hat{x}}
B_{0} \tanh [(y \pm L_{y}/4)/w_{0}]$ with $w_{0} = 2 d_{i},$
in a double tearing mode configuration with periodic boundary
conditions in all directions.  The ions are initially stationary and
initial pressure balance is enforced by a non-uniform density.  For
simplicity, we treat an isothermal plasma.  A coherent perturbation to
induce reconnection is seeded over the equilibrium magnetic field.
The resistivity $\eta$ is constant and uniform.  We use small fourth-order 
dissipation, $\propto \eta_{4} \nabla^{4}$ with $\eta_{4} = 2 \times 
10^{-5}$, in all of the equations to damp noise at the grid scale.
\begin{figure}
\includegraphics[width=3.4in]{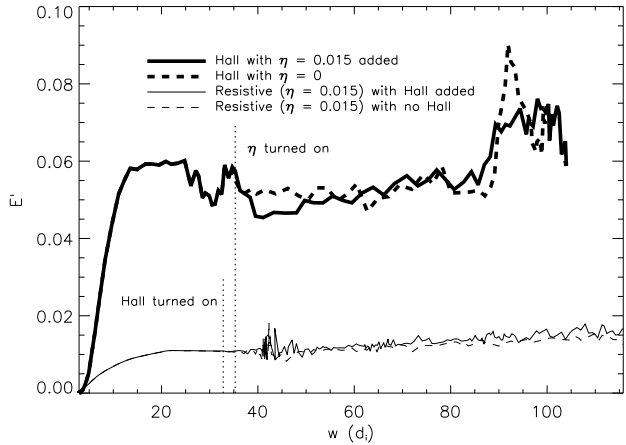}
\caption{\label{evsw} Normalized reconnection rate, $E^{\prime}$, as a 
function of island width, $w$, for the two sets of simulations described 
in the text.  The vertical dotted lines show when the added effects were 
enabled.  Note that the final parameters of the two solid line simulations 
are identical.}
\end{figure}

The computational domain must be chosen large enough to have a discernible
separation of scales between the SP and Hall reconnection rates, but with 
high enough resolution to distinguish the electron inertial scale.  We find 
that a computational domain of $L_{x} \times L_{y} = 409.6 d_{i} \times 204.8 
d_{i},$ with a resolution of $\Delta x = \Delta y = 0.1 d_{i}$ and an 
electron to ion mass ratio of $m_{e} = m_{i}/ 25$ ({\it i.e.,} $d_{e} = 0.2 
d_{i}$), is sufficient.  Since the rate of Hall reconnection is insensitive 
to the electron mass \cite{Shay98b,Hesse99,Birn01}, we do not expect our 
results to depend on our particular choice of $m_{e}$.  For this 
computational domain, we can estimate $\eta_{sf}$ and $\eta_{fs}$.  
In evaluating Eq.~(\ref{etasf}), we use $L \sim L_{x} / 4 = 102.4 d_{i}$.  
Normalizing lengths to $d_{i}$ and velocities to $c_{A0} = B_{0} / \sqrt{4 
\pi n_{0} m_{i}}$, where $n_{0}$ is the initial density far from the sheet, 
we obtain
\[
\eta_{sf}^{\prime} \equiv \eta_{sf} \frac{c^{2}} {4 \pi c_{A0} d_{i}} 
\sim \frac{d_{i}}{L} \sim 0.01.  
\]
To evaluate Eq.~(\ref{etafs}), we use the value of $B \sim 0.3 B_{0}$ 
upstream of the electron current layer measured in the simulations to 
evaluate $c_{A}$, so
\[ 
\eta_{fs}^{\prime} \equiv \eta_{fs} \frac{c^{2}} {4 \pi c_{A0} d_{i}} 
\sim 0.03.  
\]
A larger system would produce a greater separation between
$\eta_{sf}^{\prime}$ and $\eta_{fs}^{\prime}$ and would be closer to
the parameters of real systems but would be more computationally
challenging.

To demonstrate bistability of reconnection with a resistivity in the
intermediate region $\eta_{sf}^{\prime} < \eta^{\prime} <
\eta_{fs}^{\prime}$, we perform two related sets of simulations.
First, we show that a system undergoing Hall reconnection with a
resistivity below $\eta_{fs}^{\prime}$ continues Hall reconnection for
any value of resistivity below this value. We start with a benchmark
collisionless ($\eta^{\prime} = 0$) Hall-MHD simulation that is run
from $t = 0$ until the rate of reconnection is steady.  The
normalized reconnection rate $E^{\prime} = c E / B_{0} c_{A0}$ is
shown as a function of island width $w$ as the thick solid line in
Fig.~\ref{evsw}.  The reconnection rate is calculated as the time rate
of change of magnetic flux between the X-line and O-line.  The rate of 
reconnection jumps to $E^{\prime} \sim 0.06$ by the time the island width 
is $10 d_{i}$, after which it is remains steady.  When $w \sim 35 d_{i}$, 
we enable a resistivity of $\eta^{\prime} = 0.015$ (which lies between
the predicted values of $\eta_{sf}^{\prime}$ and $\eta_{fs}^{\prime}$)
and continue the simulation until most of the available magnetic flux
has been reconnected.  For comparison, the thick dashed line shows the
reconnection rate when we maintain $\eta^{\prime} = 0$.
Clearly, the reconnection rate remains nearly unchanged after the
inclusion of the resistivity.
\begin{figure}
\includegraphics[width=3.4in]{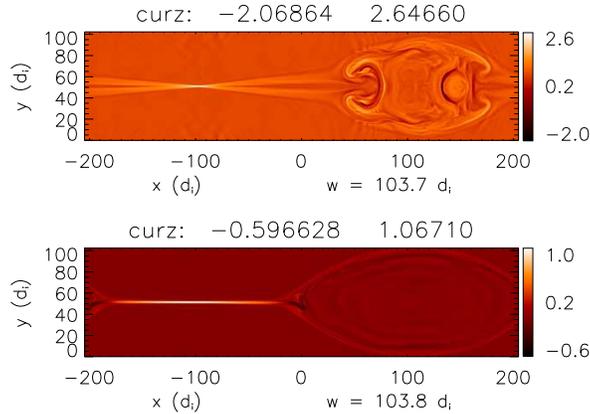}
\caption{\label{curzplot} (Color online)  Out of plane current density, 
$J_{z}$, for late times from the two solid lines of Fig.~\ref{evsw}.  The 
top plot corresponds to the thick solid line (Hall reconnection).
The bottom plot corresponds to the thin solid line (SP reconnection).}
\end{figure}

For the second set of simulations, we want to show that a system
undergoing SP reconnection continues to reconnect at the
lower rate for any value of resistivity exceeding $\eta_{sf}$. Our
computational approach is to disable the Hall and electron inertia
terms and evolve the resistive system with a resistivity that exceeds
$\eta_{sf}$. We then re-enable the Hall and electron inertia terms and
continue to advance the full equations. This benchmark simulation is 
performed with $\eta^{\prime} = 0.015$ (the same value of resistivity 
as in the run shown in the thick solid line in Fig.~\ref{evsw}), and the
reconnection rate is again plotted in Fig.~\ref{evsw} as the thin
solid line. The reconnection rate remains stationary with
$E^{\prime} \sim 0.01,$ a factor of six slower than the Hall case even
with the Hall and electron inertia terms enabled. For comparison, the
thin dashed line in Fig.~\ref{evsw} shows the reconnection rate for a
system in which the Hall term is not enabled. Thus, the Hall and the
electron-inertial terms do not impact the rate of SP reconnection for 
these parameters.

The out of plane current density, $J_{z}$, is shown at late time in
Fig.~\ref{curzplot} for the runs corresponding to the two solid curves
in Fig.~\ref{evsw}.  The top plot corresponds to the thick solid
curve.  The current sheet is short and opens wide, as is expected in
Hall reconnection
\cite{Shay99,Horiuchi97,Pritchett01,Kuznetsova01,Hesse01b,Porcelli02}.
The bottom plot corresponds to the thin solid curve.  The current
sheet is long and thin as is expected from the SP theory of
resistive reconnection \cite{Biskamp86,Uzdensky2000,Jemella04}.  Since
the same equations govern the two sets of data and the value of the
resistivity is the same, we conclude that the system is bistable.
\begin{figure}
\includegraphics[width=3.2in]{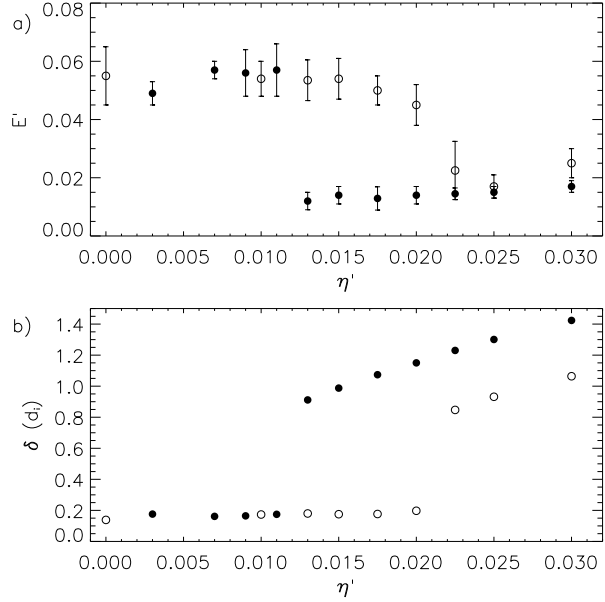}
\caption{\label{hystcurve} (a) Steady state normalized reconnection rate, 
$E^{\prime}$, as a function of normalized resistivity, $\eta^{\prime}$ for
runs analogous to those in Fig.~\ref{evsw} as described in the text.  (b) 
Current sheet width, $\delta$, as a function of $\eta^{\prime}$ for the 
simulations in (a).}
\end{figure}

To complete the mapping of the two reconnection solutions, we vary the
resistivities of the benchmark Hall and SP reconnection solutions of
Fig.~\ref{evsw}. For the case of Hall reconnection, corresponding to
the thick solid line in Fig.~\ref{evsw}, we change $\eta^{\prime}$
from 0.0 to 0.010, 0.013, 0.015, 0.0175, 0.020, 0.0225, 0.025 and 0.030 
when $w \sim 35 d_{i}$.  For the case of SP reconnection, corresponding to 
the thin solid line in Fig.~\ref{evsw}, we change $\eta^{\prime}$ from
$0.015$ to 0.003, 0.007, 0.009, 0.011, 0.013, 0.0175, 0.020, 0.0225, 0.025 and 
0.030 when $w \sim 50 d_{i}$ (after the Hall and electron-inertial terms 
have been re-enabled).  The asymptotic reconnection rate is computed as 
the time averaged reconnection rate once transients have died away.

The results are plotted in Fig.~\ref{hystcurve}(a), with the states
starting from Hall reconnection plotted as open circles and the states 
starting from SP plotted as closed circles.  The closed circles reveal 
that the disappearance of the SP solution occurs abruptly, with 
$\eta_{sf}^{\prime}$ between 0.011 and 0.013.  The open circles reveal
the disappearance of the Hall reconnection configuration, with 
$\eta_{fs}^{\prime}$ between 0.020 and 0.0225.  The error bars are due 
to random fluctuations in the reconnection rate.  The plot is 
reminiscent of what one would expect of a bifurcation diagram for a system 
with a cusp catastrophe.
\begin{figure}
\includegraphics[width=3.2in]{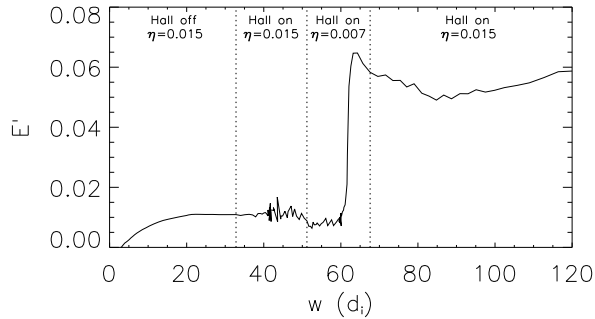}
\caption{\label{twosolns} Normalized reconnection rate, $E^{\prime}$,
as a function of island width, $w$, for the simulation which is
started at $\eta^{\prime} = 0.015$, reduced to 0.007, then increased
back to 0.015.}
\end{figure}

Thus, the numerical simulations confirm that magnetic reconnection is
bistable over a range of resistivity consistent with the scaling law
predictions of $\eta_{sf}^{\prime} \sim 0.01$ and $\eta_{fs}^{\prime}
\sim 0.03$.  The asymptotic steady state current sheet width $\delta$, 
calculated as the half width at half maximum of $J_{z}(y)$ at the X-line, 
is plotted in Fig.~\ref{hystcurve}(b) for each of the runs.  As predicted 
by Eq.~(\ref{sfcond}), the steady state SP current sheet width $\delta$ is 
of order $d_{i}$ when the resistive reconnection solution ceases to exist, 
as is shown by the closed circles of Fig.~\ref{hystcurve}(b).

We emphasize that the results presented in Fig.~\ref{hystcurve},
though generated by a specific numerical procedure, are not sensitive
to the details of this procedure. To demonstrate this, we show that the
key feature of Fig.~\ref{hystcurve}, the boundary where the slow
reconnection solution disappears can be reproduced through a
hysteresis-like procedure: in the simulation corresponding to the thin
solid line in Fig.~\ref{evsw}, we first lower the resistivity from
$\eta^{\prime} = 0.015$ to $\eta^{\prime} = 0.007$ (when the island
width is about $w \sim 50 d_{i}$).  As expected from
Fig.~\ref{hystcurve} and shown in Fig.~\ref{twosolns}, the transition
from SP to Hall reconnection occurs.  We then raise the
resistivity back to $\eta^{\prime} = 0.015$ (the original value) when
the island width is about $w \sim 68 d_{i}$.  As can be seen in
Fig.~\ref{twosolns}, fast reconnection continues, showing that the
system can be in either of two steady states for the same set of
parameters.

Having verified that reconnection is bistable with Hall MHD
simulations, we return to the onset problem.  We can compare the
critical resistivity (or temperature, using the classical Spitzer
formula) with observations of onset in physical systems such as solar
eruptions and sawtooth crashes.  For solar flares, $n \sim 10^{10}
\;{\rm cm}^{-3}, L \sim 10^{9} \;{\rm cm}$ and $B \sim 100 \;{\rm G}$
\cite{Miller97}, so from Eq.~(\ref{etasf}), $\eta_{sf} \sim 10^{-16}
\;{\rm s}$ in cgs units, corresponding to a temperature of $10^{2}
\;{\rm eV} \sim 10^{6} \;{\rm K}$.  This is in excellent agreement with
the coronal temperature.

For the sawtooth crash, the relevant kinetic scale is $\rho_{s}$
instead of $d_{i}$ because of the presence of a large guide field.  If
we assume that Eq.~(\ref{sfcond}) can be carried over to the guide
field case, we can make a comparison for sawtooth crash onset.
Typical parameters for sawteeth in the DIII-D tokamak \cite{Lazarus05}
are $B_{\varphi} \sim 2 \;{\rm T}, T_{e} \sim 2.0 \;{\rm keV}, r_{s}
\sim 20 \;{\rm cm}, n \sim 10^{14} \;{\rm cm}^{-3}$ and $Z_{{\rm eff}}
\sim 2,$ and $L \sim r_{s} \theta$, where $\theta \sim 60\,^{\circ}$
is the angular extent of the current layer \cite{Zakharov93}.  For
bean-shaped flux surfaces, the helical field strength in the plasma
core is $B \sim 100 \;{\rm G}$, so using Eq.~(\ref{etasf}) with
$\rho_{s}$ in place of $d_{i}$ yields $\eta_{sf} \sim 10^{-16} \;{\rm
s},$ corresponding to $T \sim 2 \times 10^{2} \;{\rm eV}$.  This
temperature is an order of magnitude too small.  However, the
inclusion of diamagnetic effects \cite{Levinton94}, which are known to
slow reconnection, should improve agreement. In future work we will
explore whether Eq.~(\ref{sfcond}) does hold in the presence of a
guide field, for which $\rho_{s}$ is the relevant kinetic length
scale.

The effect of collisionality on the reconnection rate was
recently explored in the Magnetic Reconnection Experiment (MRX)
\cite{Trintchouk03}.  A sharp increase in the reconnection rate
was observed at low collisionality, consistent with the qualitative
picture presented here.  Data for the current sheet width is
unavailable, so quantitative comparisons are not possible at this
time. Further, in this experiment fast reconnection has been
correlated with magnetic turbulence localized in the reconnection
layer \cite{Ji04}. Since the present simulations are limited to 2-D we can 
not address the development of this turbulence and how it might impact our
conclusions. We surmise that our conclusions will not be strongly
changed as long as the turbulence does not broaden the reconnection
layer beyond the scale length $d_i$. 

We would like to thank two anonymous referees for helpful comments
which made the presentation more clear.  This work has been supported by
NSF Grant No.~PHY-0316197 and DOE Grant Nos.~ER54197 and ER54784.
Computations were carried out at the National Energy Research
Scientific Computing Center.

%\bibliographystyle{apsrev}
%\bibliography{bib}

\end{document}